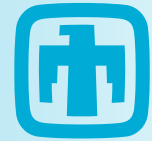

# The Synchronic Web

Thien-Nam Dinh
*Sandia National Labs*
thidinh@sandia.gov

Nicholas Pattengale
*Sandia National Labs*
ndpatte@sandia.gov

Steven Elliott
*Sandia National Labs*
selliot@sandia.gov

January 25, 2023 (last modified May 15, 2024)

**Acknowledgements** (alphabetical):

- Phillip Baxley
- Kasimir Gabert
- Curtis Johnson
- Nathan Marshall
- Ashley Mayle
- Liston Keith Purvis
- Maher Salloum
- Mara Schindelholz
- Jorge Mario Urrea

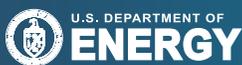
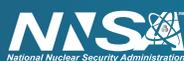

**ABSTRACT**

The Synchronic Web is a distributed network for securing data provenance on the World Wide Web. By enabling clients around the world to freely commit digital information into a single shared view of history, it provides a foundational basis of truth on which to build decentralized and scalable trust across the Internet. Its core cryptographical capability allows mutually distrusting parties to create and verify statements of the following form: "I commit to this information—and only this information—at this moment in time."

The backbone of the Synchronic Web infrastructure is a simple, small, and semantic-free blockchain that is accessible to any Internet-enabled entity. The infrastructure is maintained by a permissioned network of well-known servers, called *notaries*, and accessed by a permissionless group of clients, called *ledgers*. Through an evolving stack of flexible and composable semantic specifications, the parties cooperate to generate synchronic commitments over arbitrary data. When integrated with existing infrastructures, adapted to diverse domains, and scaled across the breadth of cyberspace, the Synchronic Web provides a ubiquitous mechanism to lock the world's data into unique points in discrete time and digital space.

This document provides a technical description of the core Synchronic Web system. The distinguishing innovation in our design—and the enabling mechanism behind the model—is the novel use of verifiable maps to place authenticated content into canonically defined locations off-chain. While concrete specifications and software implementations of the Synchronic Web continue to evolve, the information covered in the body of this document should remain stable. We aim to present this information clearly and concisely for technical non-experts to understand the essential functionality and value proposition of the network. In the interest of promoting discourse, we take some liberty in projecting the potential implications of the new model.

# CONTENTS

## LIST OF FIGURES

## LIST OF TABLES

## LIST OF ALGORITHMS

# 1. INTRODUCTION

The protection of truth in the global digital arena is among the most consequential challenges of modern times. While this effort involves many different technical and social considerations, the more specific consideration of data provenance—arguably the most basic and objective prerequisite of digital truth—is the direct subject of blockchain technology. Once limited to more specialized applications in data center management, the genesis of Bitcoin in 2009 opened the floodgates for an explosion of intense mainstream interest in this technology. Today, blockchains underpin a thriving field of creative research in sophisticated constructs such as smart contracts [31], privacy-preserving cryptography [27], layer-2 networks [25], and many others. In response, opportunistic corporations and individuals alike have co-opted the growing pool of capabilities for applications as diverse as finance and supply chains to art and identity. What has emerged is a multi-trillion-dollar industry and, in its wake, the fervor of a global socio-technical revolution. And yet, in the effort to increase trust on the Internet, the technology has fallen far short of the impact that it seems to promise. Applications that benefit all of society—rather than a handful of opportunistic investors—are few and far between. Scalability remains an insurmountable obstacle. Instead of a single source of truth and provenance, tens of thousands of platforms compete for attention[1]; the technology built to achieve consensus does not itself have a mechanism to achieve consensus.

Through the work described in this paper, we aim to revitalize the effort by returning to the core objective of blockchain technology: to propagate truth across digital networks. The guiding principle behind our approach is simplicity. Whereas most prior works have sought to create systems that are increasingly more sophisticated, performant, and expressive, we take the exact opposite approach: to do one thing, and do it well. The blockchain that underlies it is, to the best of our knowledge, the simplest, smallest, and most primitive blockchain described in technical literature. And yet, it has precisely enough functionality to order the world's information into a unified virtual ledger: a single shared view of history that we call the Synchronic Web [16].

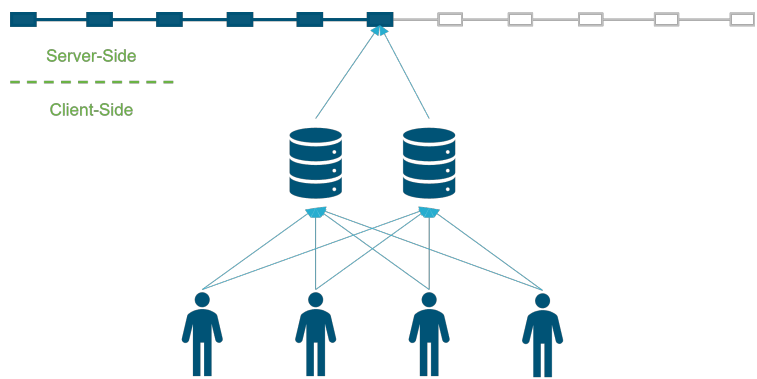

**Figure 1-1 Basic Data Architecture**

The enabling mechanism behind our design is a novel technique for committing arbitrary amounts of provenance data into fixed-sized strings. Using a group of publicly well-known, semi-trusted nodes to publish our minimalistic blockchain, the system can provide cryptographic guarantees for an arbitrary number of clients. Such clients are then able to share mathematically irrefutable statements of the following meaning: "I commit to this information—and only this information—at this moment in time". As shown in Figure 1-1, the system defers the responsibility to store and interpret data entirely to the clients. The result is a fluid, decentralized information ecosystem that achieves structure and semantic meaning through emergent, off-chain standardization rather than mandated, on-chain protocol code. To the best of our knowledge, no other system described in technical literature provides the same functionality at the same scale. The first contribution of this paper is practical: the data structures and algorithms provided are the basic blueprints of a literal blockchain network. The second contribution is conceptual: the simplest version of any complex system is often its most revelatory exemplar. By conceptualizing a system that embodies the most basic functionality of blockchain at the most extreme scale, we have a more concrete way to discuss—and make progress toward—the essential purpose of the technology as a whole. Regardless of which system or systems will ultimately achieve mass adoption, the abstractions, workflows, and semantics described in this document will be relevant to any blockchain-like network that seeks the same objective: the creation of a unified provenance framework for global cyberspace.

---
[1] https://coinmarketcap.com/

## 2. BACKGROUND

In this section, we provide a background of existing technologies that form the most important building blocks for the Synchronic Web. Note that this section only covers enabling dependencies. For a survey of technologies that provide comparable capabilities to the Synchronic Web, see Appendix A.

### 2.1. Web Technologies

Web technologies are a loosely defined set of languages, protocols, and software that enable entities to share information on "the Web" (The World Wide Web). The Web itself is a collection of information that is distributed and networked across numerous servers and clients throughout the world [7]. Its primary function is to enable automated mechanisms for finding, retrieving, and displaying static markup documents. Its key technical components include HTTP (HyperText Transfer Protocol), HTML (HyperText Markup Language), the URI (Universal Resource Identifier) scheme, and DNS (Domain Name System). The Semantic Web is a more recent concept that describes the subset of information on the Web that is networked using semantically meaningful, machine-readable vocabularies [8]. Its primary function is to enable automated mechanisms for finding, retrieving, and reasoning over facts that are digitally encoded into web documents. Key technical components include RDF (Resource Description Framework), SPARQL (SPARQL Protocol and RDF Query Language), and SHACL (Shapes Constraint Language).

As implied by the name, we consider the Synchronic Web to be a web technology for multiple reasons. First, the core infrastructure depends on protocols like HTTP for communication. Second, many of the client-side applications interact well with other web technologies like the Semantic Web. Finally, perhaps the most important type of data that it secures is information on the Web itself. One of the key reasons for the immense success of the World Wide Web is the proliferation of end-user browsers that make the Web available and useful to the general population. Similarly, one of the key variables in the success of the Synchronic Web will be its ability to integrate with end-user software like web browsers in a secure, practical, and user-friendly way.

## 2.2. Public Key Cryptography

Public key cryptography, also called asymmetric cryptography, is a method that enables mutually distrusting parties to securely exchange signed and encrypted information without the need for prior direct interaction[2]. Although public key cryptography is an active and diverse area of research, most schemes require the end-user to deal with the same two essential artifacts: a private key, which users keep in secret storage, and a public key, which users broadcast to everyone in the network [14]. Using the public key, any other user around the world can engage in secure communication with the holder of the corresponding private key. One of the most important components in public key cryptography is the PKI (Public Key Infrastructure), which is the network infrastructure responsible for hosting and sharing public keys. Today, the most successful PKI is the TLS (Transport Layer Security) PKI, which implements a hierarchical network of centralized CAs (Certificate Authorities) to secure HTTPS (HTTP Secure) communication on the Web [13]. The creation of more decentralized PKI systems continues to be an active endeavor. PGP (Pretty Good Privacy) depends on the transient nature of trust in social networks to create a "web of trust" PKI [17]. More recently, permissionless blockchains have gained traction as viable foundations for decentralized PKI systems. Many implementations of the DID (Decentralized Identifier) [29] and VC (Verifiable Credential) [28] specifications developed under the self-sovereign identity umbrella use public blockchains as public key repositories.

Public key cryptography is relevant to the Synchronic Web for two reasons. First, the nodes that operate the global infrastructure use public key cryptography to sign and verify block information. Second, clients require an existing PKI to express meaningful notions of identity. The statement "I commit to this information—and only this information—at this moment in time" is only meaningful for a given PKI. Otherwise, the statement would devolve into "someone committed this information at this moment in time". A notable feature that separates the Synchronic Web's use of public key cryptography from other blockchains is that the blockchain itself is agnostic to the PKI that the end user chooses. Wherever this paper refers to public/private keypairs, the usage is identical regardless of whether the users are interacting with TLS, PGP, Bitcoin, or any other PKI.

## 2.3. Consensus Protocols

A consensus protocol is a distributed algorithm that enables a set of potentially faulty nodes to agree on a continuously evolving shared state. A BFT (Byzantine Fault Tolerant) protocol is a consensus protocol that is resilient to "Byzantine" faults in which misbehaving nodes can act in arbitrarily malicious ways [21]. A well-established limitation of BFT protocols is that, in the best case, they can only guarantee consensus when less than one-third of the nodes are faulty. While the original paper successfully established this and other abstract properties of BFT protocols, the first concrete implementation that is widely considered to be practically useful is the appropriately named PBFT (Practical Byzantine Fault Tolerance) protocol [9]. Since the conception of PBFT, researchers have designed many other BFT variants designed to maximize various combinations of security and efficiency properties. Most early variants are "permissioned" in the sense that all participants in the consensus process are pre-approved by some authority. More recently, the rise of cryptocurrencies has catalyzed increased interest in the space of "permissionless" BFT protocols that allow any device to participate in a more decentralized manner. In the absence of an authoritative approval process, these protocols have the added challenge of implementing other ways to securely determine the influence of each node on the consensus process. For instance, proof-of-work, popularized by Bitcoin, is a permissionless consensus mechanism that uses the computational capacity of nodes[3] to inform influence [24]. As another example, proof-of-stake, adopted by Ethereum, uses the amount of "staked" cryptocurrency that a node controls to inform influence[4].

---

[2]This requirement contrasts with symmetric cryptography, which assumes that both parties have access to some previously shared secret key.

Consensus protocols are relevant to the Synchronic Web in two ways. First, the initial version of the core Synchronic Web blockchain depends on a permissioned BFT protocol to publish new blocks. Second, clients may themselves be distributed networks that can use the strong notion of discrete time offered by the Synchronic Web to augment their own consensus processes.

### 2.4. Verifiable Data Structures

A verifiable data structure is a data structure that makes use of cryptographic hashes to add immutability guarantees to the underlying information. A hash is a fixed-sized string that results from computing a cryptographic hash function over the input data. The first relevant example of a verifiable data structure is a blockchain, a list of digital objects (blocks) that each contains a payload of arbitrary data and a hash of the previous block in the list[5]. These data structures allow mutually distrusting parties to verify the integrity of an arbitrary amount of immutable data by achieving consensus over only the hash of the last block. If the number of blocks is $n$, then the size of the blockchain is $O(n)$ and the time to verify the integrity of a specific block is $O(n)$. The second relevant example of a verifiable data structure is a Merkle Tree, a binary tree in which each leaf is the hash of some piece of arbitrary data and each non-leaf is the hash of its two child nodes [23]. The root node of the Merkle Tree is, by implication, an indirect hash of all pieces of data that are hashed into its leaves. Like blockchains, Merkle Trees allow parties to verify the integrity of data by sharing only the root of the data structure. However, in exchange for the extra complexity imposed by its tree structure, the verification procedure is categorically more efficient. If the number of leaves is $n$, then the size of the Merkle Tree is $O(n)$ and the time to verify the integrity of a specific leaf is $O(\log n)$. Due to its widespread use in security applications, researchers have defined many specialized variants of Merkle Trees. Sparse Merkle Trees [2] and Patricia-Merkle Trees [31] are two widely used variants that use the location of leaf nodes to represent meaningful metadata.

Verifiable data structures are especially relevant to this paper because the Synchronic Web is, in a technically accurate sense, nothing more than a massively distributed data structure. More precisely, it is a blockchain of Merkle Trees whose leaves consist of all data that has ever been com-mitted by a client. Furthermore, the main technical contribution of this paper is the novel application of a specialized Merkle Tree that we refer to as a verifiable map[6]. Although our implementation optimizes toward a specific use case, its key properties are comparable with Sparse Merkle Trees, Patricia-Merkle Trees, and other similar variants.

---

[3] More precisely, the capacity is the number of cryptographic hashes that it can calculate within some period.
[4] https://ethereum.org/en/developers/docs/consensus-mechanisms/pos
[5] The only object that doesn't contain a hash is the zeroth block, which is often referred to as the genesis block.

## 3. DESIGN

In this section, we describe the core capability of the Synchronic Web. Although the foundational infrastructure—a simple blockchain that secures the state of the system—should remain stable, the ecosystem around this infrastructure can evolve toward increasingly complex designs. This section is our attempt to recommend one compelling starting point. Section 3.1 introduces the core design through the lens of a concrete motivating example while Section 3.2 presents several additional technical considerations at a more general, abstract level.

### 3.1. Core Procedure

The example scenario assumes a situation in which the Synchronic Web infrastructure is widely accessible on the Web. The core blockchain is operated by a consortium of trusted public organizations that run notaries, special web servers that participate in the blockchain consensus protocol and provide the essential Synchronic Web service to clients, which we call ledgers and verifiers. Although notaries operate independently, they cooperate well enough that clients can treat the Synchronic Web as one logical notary. Our example scenario describes a three-party protocol between a generic notary and the following two clients:

> **Ledger:** Alice is the founder of *The Looking-Glass*, an up-and-coming digital media news outlet for high-impact publications ranging from investigative journalism to coverage of geopolitical events. Notably, the organization is known for its excellent use of technology to drive user engagement: creative social media integrations, a top-of-the-line data model for recommendations, and sophisticated visuals for sharing real-time statistical analysis with consumers. As a writer-turned-entrepreneur, Alice is particularly attuned to the importance of intellectual ownership and makes sure that *The Looking-Glass* signs every publication with its well-known cryptographic identity. As a result, when she learned about a software development kit that could leverage her existing signing key to provide strong notions of data provenance, she immediately directs her team to integrate it into their data management workflow.

> **Verifier:** Bob is an average Internet user who occasionally frequents news outlets like *The Looking-Glass*. Although Bob is far from a privacy or security activist, he harbors a healthy skepticism toward online media and the platforms that proliferate it. As a result, when a friend told him about a light-weight browser extension that can automatically verify the authenticity of data on participating websites, he mustered up just enough motivation to click "install".

---

[6] We borrow this term from a more general description founded in existing literature [2].

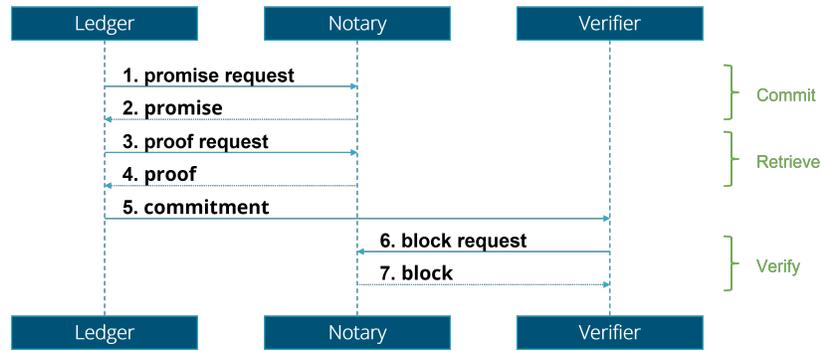

**Figure 3-1 Canonical Interaction Sequence**

Figure 3-1 illustrates the high-level procedure that takes place for each relevant piece of content shared between Alice and Bob. For brevity, we refer to *The Looking Glass's* data management system as simply "Alice" and Bob's Synchronic Web browser extension as simply "Bob". First, Alice commits content to the Synchronic Web. Once committed, she requests a proof that makes it accessible to Bob from her organization's public website. Finally, Bob compares the proof against the root of the Synchronic Web blockchain to verify its integrity. The following sections describe each step of the protocol in greater technical detail. Throughout the section, we distinguish between *local* variables, which the ledger uses to commit data into its personal state of history, with *global* variables, which the notary uses to combine local states into a single shared state.

### 3.1.1. Commit

In the first stage, Alice commits her global content to the Synchronic Web, which she computes in the following way. First, Alice creates a map in which each value is the hash of a publication, which we call the local content, and each key is a hash of a unique identifier for the publication, which we call the local path. A sensible path to use for each publication is the URL where Bob can find it on *The Looking Glass's* website. Next, Alice uses Algorithm 1 to compute a special Merkle Tree over the data structure, effectively turning it into a local verifiable map with a local root. This data structure is the global content that she must commit to the Synchronic Web. Next, Alice declares a persistent global path to which she will commit her global content. This global path should be a well-known string that everyone else in her industry uses to publish their global content. She now has all the inputs necessary to compute the following outputs:

$$\text{key}_{\text{global}} \leftarrow \texttt{Hash}(\texttt{Sign}(\texttt{Encode}(\text{index}, \text{public-key}, \text{path}_{\text{global}}), \text{secret-key})) \qquad (1)$$

$$\text{value}_{\text{global}} \leftarrow \texttt{Hash}(\texttt{Encode}(\text{root}_{\text{local}})) \qquad (2)$$

This global key-value pair defines the full state of *The Looking Glass* at the time associated with the target index within the semantic context implied by the global path. Alice sends these outputs, along with the target index, to the notary.

**Algorithm 1 Get Tree.** Accepts a standard key-value map and returns a Merkle Tree for generating inclusion proofs over the provided map.

---

1: **function** GETTREE(map)
2:     **function** Recurse(items, depth)
3:         **if** items.length $= 0$ **then**
4:             **return** Node($\varnothing, [\varnothing, \varnothing]$)
5:         **else if** items.length $= 1$ **then**
6:             **return** Node(items[1], $[\varnothing, \varnothing]$)
7:         zeros, ones $\leftarrow$ [], []
8:         **for each** item $\in$ items **do**
9:             **if** Binary(item[0])[depth] $= 0$ **then**
10:                zeros.append(item)
11:             **else**
12:                ones.append(item)
13:         left, right $\leftarrow$ Recurse(zeros, depth $+1$), Recurse(ones, depth $+1$)
14:         **return** Node(Hash(left[0] $+$ right[0]), [left, right])
15:     **return** Recurse([[key, Hash(key $+$ value)] **for each** key, value $\in$ map], 0)

---

The notary accepts Alice's request along with requests from other ledgers throughout the world. Eventually, enough time elapses that it is time to aggregate all requests associated with the specified index and lock them into a new block in the blockchain. To do so, the notary passes the map of all global key-value pairs it has received into Algorithm 1 to create the global verifiable map. Finally, together with all the other nodes in the global consensus protocol, it publishes the global root of this verifiable map to the payload of a new block with the corresponding index, thus locking every piece of data described in this section—from each opaque global key-value pair to the local paths and content used to compute them—into the new immutable state of history.

### 3.1.2. Retrieve

Now that Alice has successfully committed the contents of *The Looking-Glass*, she still needs to create a proof of inclusion that traces the global root down to each piece of local content. To do so, she simply sends the global key that she computed in Equation 1 and the corresponding index to the notary. Using Algorithm 2, the notary parses the global verifiable map to complete the corresponding global proof and send it back to Alice. Alice can then combine this global proof along with a local proof, computed by executing Algorithm 2 over her local verifiable map, to create a complete commitment over any piece of local content that she originally committed in Section 3.1.1. Alice can now publish the commitment on her website for all to see.

### 3.1.3. Verify

Somewhere across the Internet, Bob opens an article on *The Looking-Glass*. Looking through the website's metadata, he (more precisely, his browser extension) finds a well-formed Synchronic Web

**Algorithm 2 Get Proof.** Accepts a Merkle Tree and key and returns an inclusion proof that can trace the list of intermediate nodes connecting the key to the tree root.

```
1: function GETPROOF(tree, key)
2:     node ← tree.root
3:     proof ← []
4:     for each bit ∈ Binary(key) do
5:         if node.children = ∅ then
6:             break
7:         proof.append(node.children[(bit + 1) mod 2].digest)
8:         node ← node.children[bit mod 2]
9:     return proof
```

commitment. First, he uses Algorithm 3 to infer the local root used in Equation 2 from the provided local path and local content. Next, he uses Alice's well-known public key to verify Equation 1. Finally, he uses Algorithm 3 to infer the global root of the Synchronic Web. If this global root matches the blockchain root that it receives from the notary, then he has clear, irrefutable evidence for the validity of Alice's commitment.

**Algorithm 3 Get Root.** Accepts a key, value, and inclusion proof and returns the implied root of the Merkle Tree to which the key belongs.

```
1: function GETROOT(key, value, proof)
2:     step ← Hash(key + value)
3:     for each i ∈ [1...proof.length] do
4:         if Binary(key)[proof.length − i] = 0 then
5:             step ← Hash(step + proof[proof.length − i])
6:         else
7:             step ← Hash(proof[proof.length − i] + step)
8:     return step
```

Figure 3-2 provides a visual of the full global verifiable map that contains Alice's commitment. At this point, it is worth reiterating the semantic implications of the cryptographic guarantees.

- For any given index, Alice can create only a single valid global key according to Equation 1.

- For any given local key, the local verifiable map rooted in Equation 2 can only hold a single local value corresponding to the local content.

- For any given global key, the global verifiable map rooted in the Synchronic Web block can only hold a single global value corresponding to the local root.

In other words: for any given index, global key, and local key, Alice can commit to one—and only one—piece of content. Regardless of where the article may later be republished (e.g., in Bob's social media post, Carol's inbox, or Dan's congressional briefing), the presentation of its corresponding commitment permanently secures its provenance to a single point in time and space on the Synchronic Web.

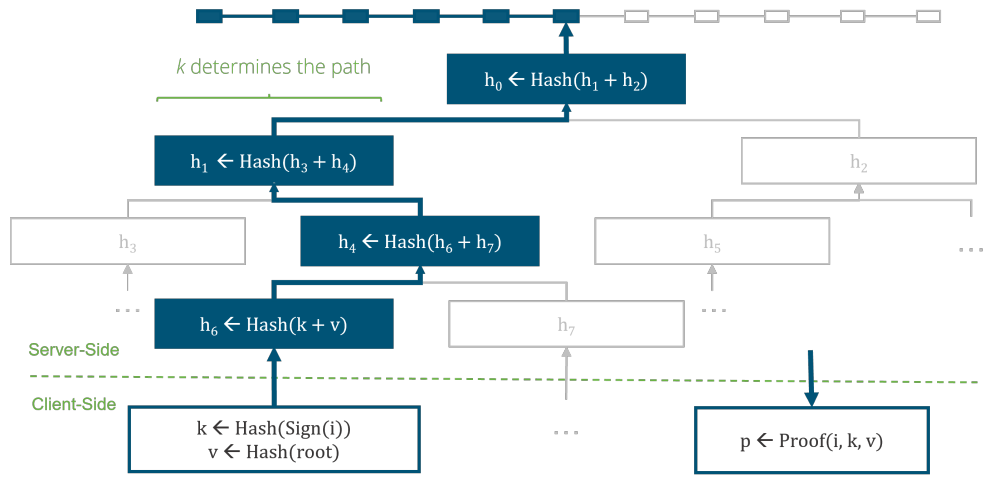

**Figure 3-2 Verifiable Map Tree Structure**

## 3.2. Technical Considerations

In this section, we next discuss some of the idiosyncratic technical details at play behind the procedure of Section 3.1.

### 3.2.1. Scalability

The distinguishing value proposition of the Synchronic Web is that the notary network can theoretically provide its core functionality—committing ledger-submitted information to the blockchain—at arbitrary scale. There are two main components to scalability.

**Throughput:** Without compromising the core guarantees of the final data structure, notaries can generate global verifiable maps in an arbitrarily distributed fashion. This procedure entails parallelizing the recursive calls in Algorithm 1 such that $n$ workers could generate a verifiable map in $O(\log_T n)$ time[7]. Similarly, each ledger can further parallelize the creation of its local verifiable map, ensuring that anyone with sufficient resources can commit data at arbitrary granularity with $O(\log_2 n)$ commitment sizes.

**Storage:** After notaries create a verifiable map and publish its root to the blockchain, it must retain this data for some well-established, fixed period to allow for ledgers to retrieve their requested proofs. However, after that time has passed, the notary can discard most of the verifiable map and free up space for new requests. The root is the only piece of information that the network retains at a global level, which it stores forever in the incrementally growing blockchain. Thus, the Synchronic Web is scalable with respect to storage because the burden of retaining content and commitments rests entirely with the ledger. Ledgers are free to store the data using arbitrarily complex datastores that support features such as replication and federation. Regardless of the chosen datastore, ledgers are also able to arbitrarily split off or merge historical commitments from other ledgers that share the same semantic data model[8].

---

[7]Here, the base of the logarithm T is the throughput of any single worker.

An important consequence of the inherent scalability provided by the core design is to limit the complexity of client integrations. From a resource perspective, permissioned access control lists and permissionless cryptocurrencies on conventional blockchains are nothing more than mechanisms to allocate scarce storage space. Since the Synchronic Web promises no such storage, the infrastructure needs no such complexity. However, like many web services, it does require basic mitigation against denial-of-service attacks.

### *3.2.2. Synchronization*

In many situations, data provenance is only useful if the ledger can show changes in the same content over time. For instance, in the case of the *The Looking-Glass* example from Section 3.1, it is not enough for Bob to only have access to Alice's ledger at time $i$, since Alice might be tempted to present a completely different view to Carol at time $i+1$. To prove that she hasn't done so, Alice can embed an incrementing sequence number into the content of each commitment that she creates in a well-standardized format. This allows Bob to view all commitments within some time window and be sure that he possesses an unbroken chain of content versions. In normal operating situations, the commitment sequence number will increase in lock-step with the blockchain index. However, if a ledger experiences network downtime, then it will be unable to commit to some indices. The job of the verifier is to make sure that these gaps are not so egregious that they could hide a maliciously created fork of the content. The test for this misbehavior is straightforward: for some time window of interest, the set of sequentially contiguous commitments is valid if and only if they cover more than half of the expected indices.

Due to the need for ledgers to make regular commitments, an important decision they must make is to determine the frequency at which they synchronize with the notary—a decision largely determined by the domain and use case. To accommodate various needs, we introduce the concept of periodicity. Periodicity, in the context of the Synchronic Web, determines the commitment frequency of each ledger according to the following formula:

$$\text{period} = \frac{1}{\text{frequency}} = 2^{\text{periodicity}} \tag{3}$$

In other words, a ledger that signals a periodicity of $0$ commits to every block, $1$ commits to every $2$ blocks, $3$ commits to every $4$ blocks, and so on. Importantly, the index of the submitted blocks *must* be divisible by the period. The benefit of this scheme is that it allows lower-periodicity clients to interoperate with high-periodicity clients since all periods are ultimately powers of $2$. In addition, certain niche use cases may require a more granular notion of time than the global Synchronic Web notary provides. Entities that require this level of granularity might consider setting up high-frequency notaries that commit aggregate checkpoints to the global network. For these use cases, the concept of periodicity extends naturally to allow negative numbers for defining block intervals of less than $1$.

---

[8]The process of sharing data between ledgers in the Synchronic Web is analogous to the process of sharing data between shards in conventional blockchains.

### 3.2.3. Availability

The primary challenge in the design of the notary infrastructure is to guarantee censorship-resistant service availability. Censorship resistance directly implies the protection of privacy for ledgers. Since communications between ledgers and notaries involve only cryptographic hashes, the primary privacy vulnerability is the communication channel itself. A privacy-conscientious notary infrastructure implementation could therefore offer basic routing misdirection to prevent any one entity from censoring specific ledgers. In addition, particularly privacy-conscientious ledgers could bolster the communication privacy of their communication channels by using independent VPNs (Virtual Private Networks) or anonymous routing networks [15].

The more difficult design challenge is to guarantee the availability of the notary service. To ensure that notaries are accountable for their promises, ledgers should retain signed copies of each promise message it receives in Section 3.1.1. If ledgers retain these messages properly, there are two levels of guarantee that they could theoretically obtain from the network.

**Strong:** The network guarantees that for any promise, the corresponding commitment will be included in the promised block.

**Weak:** The network guarantees that for any notary that does not include a promised commitment in the agreed-upon block, there exists a protocol that will remove the faulty notary from the consensus protocol within $b$ blocks.

Strong availability places a burden on the network of notaries to securely share responsibility for commit requests. Weak availability places a burden on ledgers to recover from situations in which they deal with faulty notaries. However, well-designed ledgers should already include mechanisms to recover from connectivity failures which, from the perspective of the ledger, is essentially indistinguishable from a temporarily faulty notary. The design of an optimal availability mechanism is a high priority for future research. A more formal security model to contextualize this work is provided in Appendix B.

### 3.2.4. Extensibility

The ethos of the Synchronic Web emphasizes standards-driven extensibility. Although communities of users may alter or grow any aspect of the design, the following areas are immediate candidates for more granular domain-specific extensions.

**Resolution:** Resolution is the process of encoding content that can be hashed into a commitment. Users can write new resolver specifications to integrate data from increasingly diverse sources.

**Authentication:** Authentication is the process of using public key cryptography to link a commitment with a digital identity. Users can write new authenticator specifications to integrate identities from increasingly diverse PKIs.

**Linking:** Linking is the process of connecting multiple commitments to assert new semantically meaningful claims. The use of sequence numbers described in Section 3.2.2 is an example of linking. Connecting co-located content (content validated by different authenticators at the same index, global path, and local path) is another example. Users can write new linker specifications to define increasingly diverse use cases for the Synchronic Web.

## 4. USAGE

This section categorizes the use cases of the Synchronic Web. Whereas Section 3 focuses on a single use case, the value of the Synchronic Web is that it can provide a foundation for all use cases and an unbounded number of clients—all under the same framework. For instance, it may secure content as diverse as DNS logs, tax receipts, or visual art. It may support functionality as specialized as punctual proof-of-work protocols[9] or un-forkable proof-of-stake[10]. The setting could conceivably vary tremendously as well, ranging from open-web publications to highly sensitive classified or proprietary networks. Arguably, the more interesting question, then, is not *what* data benefits from provenance, but *why* data benefits from provenance. Figure 4-1 presents a basic delineation based on the relationship model between the ledger and the verifier. Using this framework, we describe each category in greater detail.

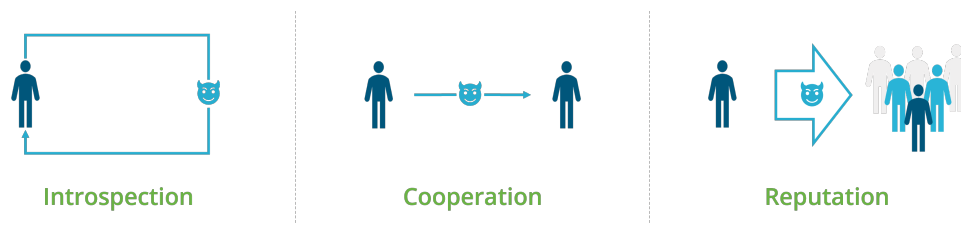

Introspection      Cooperation      Reputation

**Figure 4-1 Usage Relationship Models**

### 4.1. Introspection

In introspective use cases, entities create Synchronic Web commitments that they later verify themselves. An instructive example of introspection is cybersecurity logging in enterprise networks. In modern networks, it is good practice to retain logs for critical activities such as logins, command line inputs, and certain network communications. By monitoring the appropriate system content, the system can help incident responders discover indicators of compromise in the event of a security breach. However, cybersecurity logging is ineffective if the attacker can breach the logs themselves and remove evidence of malicious activity. For most systems today, the best technique to prevent covert log manipulation is to copy the logs off-premises. With the Synchronic Web, enterprise systems can locally secure cybersecurity logs in a way that, once committed, is prohibitively difficult for adversaries to alter without clear evidence of tampering. This measure would increase the trust that an organization has in its information systems.

---

[9] In existing proof-of-work blockchains like Bitcoin, the miner who first achieves the mining target publishes a new block probabilistically every $n$ seconds. With the Synchronic Web, it is possible to implement a proof-of-work variant such that the miner who gets closest to the mining target publishes a new block precisely every $n$ seconds.

[10] In existing proof-of-stake blockchains like Ethereum, the network is vulnerable to long-range "nothing-at-stake" attacks in which a malicious coalition of validators who controlled a staking majority at some point in the past can produce fraudulent versions of history. With the Synchronic Web, it is possible to force validators to commit to a specific version of history such that clients can later distinguish valid from invalid forks of the same ledger.

More broadly, any entity that operates a complex information system could benefit from secure data provenance. For instance, online banking and electronic voting platforms are two examples of applications in which secure provenance is particularly desirable. Even more broadly, the concept of introspection can apply to social networks. In the same way that computer breaches misappropriate code to exert control over computer systems, disinformation campaigns misappropriate natural language to exert control over social systems. The relevant content to monitor for evidence of disinformation include news articles, videos, and social media messages. In any introspective use case, the basic benefit of the Synchronic Web is to increase the confidence level at which analysts can study historical and ongoing misbehavior. However, it is likely that in many cases, the mere act of information retention—and the threat of future detection—is enough to deter certain types of threats.

### 4.2. Cooperation

In cooperative use cases, entities create Synchronic Web commitments that interacting counterparties later verify. An instructive example of cooperation is treaty verification between sovereign states. The exchange of trustworthy information is an important aspect of diplomacy in the international arena. In processes like inventory reporting under the Chemical Weapons Convention or sensory data under the Comprehensive Nuclear Test Ban Treaty, better data integrity can improve trust and reduce the chance of conflict[11]. In some cases, the treaty participants themselves, rather than some external attacker, may have an incentive to doctor self-reported data. Such misbehavior can be difficult to detect and prove in the arbitration process. By requiring participants to commit all relevant data to the Synchronic Web, treaty counterparties could achieve greater confidence in the degree of cooperation.

More broadly, any entity that is involved in a contractual relationship could benefit from secure data provenance[12]. For instance, a product supply chain is a complex series of contractual relationships between commercial industries. By committing data to the Synchronic Web, participants in the supply chain can decrease liability and increase accountability of product inventories. Whereas treaty verification and supply chains typically involve peer-to-peer relationships, cooperation could also involve hierarchical relationships, as in the case of government regulations on private businesses. Many financial and safety regulations, for instance, mandate the retention of various kinds of documentation. By mandating integration with the Synchronic Web, regulatory bodies could improve the quality of routine documentation and, consequently, the efficacy of related auditing processes.

---

[11]Under realist international relation theories, a major cause of warfare is the communication asymmetry between states in which even a misperception of military mobilization in one country leads to a disproportionate escalation in others. These scenarios are referred to as "security dilemmas."

[12]In this section, a contract can be a legally binding agreement, an informal mutual understanding, or anything in-between.

## 4.3. Reputation

In reputational use cases, entities create Synchronic Web commitments that the general community later verifies. An instructive example of reputation is scientific publications in academia. In the collective endeavor to make scientific progress, the individual incentive for any given scientist is to show that they were the first to discover some new piece of knowledge. Traditional academia depends primarily on social mechanisms to assign credit for discoveries: publications to prestigious conferences, recognition by influential colleagues, etc. The Synchronic Web provides a more standardized way for scientists to assert provenance over work. In addition to timestamping the final publication, scientists could also leverage a persistent cryptographic identity to establish the provenance of supplementary content like raw data sets or lab notes. Ideally, this capability would be integrated into end-user software like document processors and electronic lab notebooks to the point where ubiquitous provenance for scientific discoveries—and due credit for the innovative researchers behind them—becomes the norm.

More broadly, any entity that desires to establish its reputation in a distributed social context could benefit from secure data provenance. In a similar vein as the more specific academic example, the following endeavors all entail a strong interest in reputation.

> **Discovery:** Scientists, philosophers, and other discoverers build reputations by formulating ideas that add new insight to human understanding. They can prove credit for their work by committing content like raw data, personal notes, and photographic evidence of discoveries.
>
> **Invention:** Engineers, entrepreneurs, and other inventors build reputations by identifying approaches that add new efficacy to human endeavors. They can prove credit for their work by committing content like blueprints, computer-aided design models, and test results.
>
> **Art:** Writers, musicians, and other artists build reputations by creating experiences that add new meaning to the human experience. They can prove credit for their work by committing content like recordings, sketchbooks, and finished works.

In many use cases, the simple act of timestamping an original piece of work next to a first and last name is enough to establish due credit. In other use cases, there is value in using cryptographic identities to enforce Synchronic Web commitments that allow one—and only one—version of relevant content. For instance, the "replication crisis" in scientific publications highlights the importance of enforcing reporting standards in the research process [11]. One of the most important efforts that have emerged from this crisis is the effort to encourage the preregistration of experiment hypotheses and plans to ensure the statistical significance of the subsequent results. Researchers can use the Synchronic Web to prove compliance by recording experimental preregistration, results, and analysis in the same ledger. Other domains could benefit from the same capability. For instance, professionals who frequently make predictions can write them in their ledgers to show that they commit to one—and only one—cohesive set of thoughts. Examples include economic analysts, political advisors, and sports pundits. Independently of the domain, the Synchronic Web provides an important capability in the task of securely scaling reputation in the digital age.

# 5. OUTLOOK

This section speculates on the outlook for the Synchronic Web in the context of the overarching task of achieving widespread adoption and usage.

## 5.1. Technical Development

The first component we discuss is technical development. Although stability is a fundamental design goal of the specific infrastructure described in this document, the adjacent technology stack that it brings together can—and will—continue to evolve. To promote productive and controlled growth, we suggest an "hourglass" model of development inspired illustrated in Figure 5-1 and inspired by popular descriptions of Internet architecture[13]. The following two sections describe outlooks for the application and infrastructural layers in more detail.

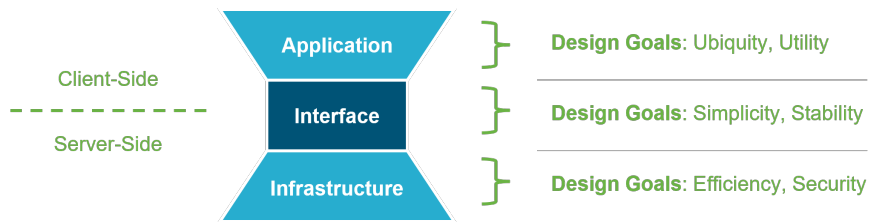

**Figure 5-1 Hourglass Architecture Model**

### 5.1.1. Server-Side

Server-side development seeks to improve notary implementations, consensus protocols, and other infrastructural components to increase the efficiency and security of the Synchronic Web. Notable subcategories for infrastructural development are as follows:

**Consensus:** Consensus in the Synchronic Web infrastructure is the process by which authorized notaries agree on new blocks. Whereas traditional BFT protocols solve consensus over purely on-chain data, the Synchronic Web introduces the more challenging problem of consensus over off-chain requests that contribute only to the preimage of the block payload. Although Section 3.2.3 discusses some properties of acceptable solutions, additional research is required to achieve better concrete guarantees of availability. An ideal mechanism would guarantee that the network fulfills ledger commitment requests with the same assurance that traditional BFT protocols provide for on-chain transaction requests.

**Authorization:** Authorization in the Synchronic Web infrastructure is the process by which society agrees on the list of organizations that can participate in the global consensus process. Since traditional permissioned protocols assume access to some list of authorized nodes, the method for selecting trustworthy participants in the first place remains an open socio-technical challenge. Whereas cryptocurrencies solve the problem through market-based mechanisms (e.g., proof-of-work, proof-of-stake, etc.), a system like the Synchronic Web would likely benefit from a more democratic mechanism such as a web-of-trust or a direct voting system. The ideal authorization mechanism would dynamically evaluate and authorize only the most well-known entities according to democratic consensus.

---
[13]https://cacm.acm.org/magazines/2019/7/237714-on-the-hourglass-model/fulltext

Like in any complex information system, the need to optimize efficiency and security often conflicts. Although it is currently possible to separately describe either an arbitrarily scalable or a fully Byzantine fault-tolerant configuration of the Synchronic Web, the existence of a configuration that achieves both properties simultaneously is a currently open research question. Regardless, by further exploring the design space, researchers should be able to achieve better trade-offs between the two dimensions to some degree.

### *5.1.2. Client-Side*

Client-side development seeks to improve ledger implementations, end-user software, and other application technology to increase the ubiquity and utility of the Synchronic Web. Notable subcategories for application development are as follows:

**Data Storage:** Development in data storage entails integrating ledgers with advanced database technologies to support more sophisticated use cases. For example, many centralized database systems offer high-availability features[14,15,16] that complement the integrity guarantees provided by the Synchronic Web. More decentralized examples include peer-to-peer datastores[17,18] that rely on content addressable storage networks like IPFS (InterPlanetary File System) for replication and censorship resistance [5]. Finally, the Synchronic Web can augment integrity guarantees provided by existing immutable database technologies[19,20,21] that offer append-only, tamper-evident logging [12]. An ideal data management integration would enable end-users to interact with Synchronic Web commitments in the context of a data model native to the database itself.

**Cryptography:** Development in cryptographic integration entails combining ledgers with advanced cryptographic primitives to enable privacy-preserving commitments. Examples of advanced constructs include implementations of zero-knowledge proofs [18], secure multi-party computation [32], and fully homeomorphic encryption [26]. Healthcare data mining, regulatory compliance, and international alliances are examples of domains where privacy-aware data sharing is desirable for individuals, corporations, and governments, respectively. An ideal cryptographic integration would enable end-users to prove and verify the validity of

---

[14] https://www.oracle.com/database/data-guard/
[15] https://mariadb.com/kb/en/orchestrator-overview/
[16] https://www.mongodb.com/docs/manual/replication/
[17] https://github.com/orbitdb/orbit-db
[18] https://gun.eco/
[19] https://immudb.io/
[20] https://www.datomic.com/
[21] https://aws.amazon.com/qldb/

Synchronic Web commitments without revealing any more information about the data than is required by the use case.

**Semantics:** Development in semantic integration entails combining ledgers with semantic web vocabularies to facilitate more secure knowledge interoperability on the open web. An example of a vocabulary that complements the core Synchronic Web functionality is the PROV-O ontology for digital provenance [22]. Example vocabularies that can improve the interoperability of ledger authenticators include DIDs [29] for asserting control of digital identities and VCs [28] for asserting statements about digital identities. Finally, examples of vocabularies that improve the interoperability of ledger content include schema.org [19] for structured search engine results and Data Catalogs [3] for FAIR (Findable, Accessible, Interoperable, and Reusable) scientific data [30]. An ideal semantic integration would allow applications that make use of these external semantic vocabularies to also prove and verify Synchronic Web commitments.

These research directions describe academically novel areas of work that would help grow the space of Synchronic Web use cases. However, perhaps the larger challenge—and opportunity—lies in the entrepreneurial development of software products that leverage these foundations for increasingly new and creative end-user applications.

### 5.2. Economic Considerations

The second component we discuss is economic motivation. Conventional blockchain networks are profit-driven. Whereas permissionless networks like Bitcoin and Ethereum seek to maximize their market capitalization, permissioned networks like Hyperledger Fabric and Quorum seek to maximize the corporate adoption of their software services. These profit-driven economic models are necessary because the core service of conventional blockchain networks—secure data storage—is fundamentally scarce. In contrast, the core service of the Synchronic Web—guarantees of data immutability—is arbitrarily scalable. The result is that the public value of obtaining the service vastly exceeds the marginal cost of providing it, making it a quintessential public good[22]. One of the many economic peculiarities of the digital age is the growing proliferation and influence of these goods. We list some relevant examples below:

**Infrastructure:** Public infrastructural goods provide foundational capabilities for a large population. DNS is a classic example of a collective Internet service. Examples of collectively developed and maintained overlay networks include Tor, IPFS, Let's Encrypt CA, and fediverse services like Mastodon that make use of the ActivityPub protocol. Finally, two examples of systems that link the digital world with the physical world are the Global Positioning System and the International Atomic Time system.

**Security:** Public security goods decrease the risk of negative consequences from adverse events for a large population. Relevant examples include natural monitoring systems such as the Global Forecast System, the Global Seismographic Network, the Pacific Tsunami Warning and Mitigation System, and Sentry, an asteroid collision monitoring system. Another example is the Comprehensive Nuclear-Test-Ban Treaty Organization's International Monitoring System, which addresses the man-made threat of nuclear weapons testing.

---

[22]This term refers to the economic concept of a non-rivalrous and non-excludable good or service that is available to all members of society.

**Information:** Public information goods persist shared knowledge for a large population. Wikipedia is the most well-known example of a community-maintained and freely available knowledge repository. The Internet Archive, primarily through its Wayback Machine, provides a digital archival service that is particularly aligned with some objectives of the Synchronic Web. Numerous other examples exist for more niche domains such as arXiv for scholarly publications and Project Gutenberg for books in the public domain.

The economic drivers behind these systems are as diverse as the capabilities themselves. However, the broad categories of investors are clear: security-minded governments, mission-oriented nonprofits, and mutually benefiting private sector consortiums. At an individual level, more modern perspectives from the digital age hint at the surprising power of aggregated intrinsic motives to drive large-scale innovations [6]. What has become clear is that the dynamics that drive progress in this arena are radically different from the market-based mechanisms that have traditionally driven the blockchain industry. It is through this paradigm the Synchronic Web aims to repurpose the astounding technological progress that private capital has purchased. Here, in the absence of scarcity, what drives design is a preference toward cooperation, accessibility, and impact. What emerges is a chance to build the world's first truly secure basis for global provenance: a single, unified structure for us to inscribe the new epochs of digital history.

# APPENDIX A. RELATED WORK

This section enumerates technologies that are related to the Synchronic Web and describes their relative advantages and disadvantages. Table A-1 provides an overview of these comparisons along three dimensions: (1) *capability* describes the variety of use cases that a technology supports, (2) *security* describes the degree to which it can guarantee availability and correctness for those use cases, and (3) *scalability* describes the number of clients for which it can support those use cases. Rather than making concrete assertions about the value of these technologies, the comparisons are only intended to highlight key differences with the Synchronic Web as of the writing of this paper.

|                                  | Capability | Security | Scalability |
|----------------------------------|------------|----------|-------------|
| **Timestamping Authorities**     | Lower      | Lower    | Similar     |
| **Blockchain Timestamping**      | Lower      | Similar  | Similar     |
| **Distributed Ledger Technology**| Higher     | Similar  | Lower       |

Table A-1 Comparison of Related Technologies

## A.1. Timestamping Authorities

TSAs (Timestamping Authorities) are centralized services that use their trusted authoritative status to provide clients with timestamped cryptographic signatures over arbitrary data. Many organizations provide TSA services for time-sensitive information like legal documents, intellectual property, and forensics[23,24,25]. The IETF (Internet Engineering Task Force) provides the primary specification for this procedure [1].

TSAs are similarly scalable relative to the Synchronic Web. However, since the system depends entirely on the existence of trusted, centralized authorities, TSAs are less secure. If an adversary compromises a TSA, then it would be able to covertly generate fraudulent timestamps for both historical and future documents. In contrast, to generate fraudulent timestamps in the Synchronic Web, the adversary would have to compromise at least two-thirds of the notaries, a feat that is not only prohibitively difficult but also highly noticeable. In addition, TSAs offer lower capability because, according to the defining standard, they do not offer the ability for clients to prove that they are only committing to one version of content data within a discrete time step. Although an organization like the IETF could extend the standard to mimic this capability, it would introduce the additional security risk that TSAs could lie about the uniqueness of client requests.

---

[23] https://digistamp.com
[24] https://forensicnotes.com
[25] https://freetsa.org

### A.2. Blockchain Timestamping

BTS (Blockchain Timestamping) services are centralized services that use public blockchains to timestamp data using cryptographic hash data structures. There are a growing number of BTSs that occupy a similar market niche as TSAs[26,27,28]. Although the specific implementations may vary, most BTSs operate by accepting requests from clients, batching requests into a Merkle Tree, and committing the root of the Merkle Tree as a transaction on some well-known blockchain such as Bitcoin.

BTSs are similarly secure and scalable relative to the Synchronic Web. However, they are less capable because, like TSAs, they do not have a mechanism to prevent clients from committing to multiple versions of the same content. The most basic form of BTS functionality is, therefore, a strict subset of Synchronic Web functionality. BTSs are often operated by smaller organizations that rely on the security of large, well-known blockchain networks like Bitcoin. These organizations also tend to use specialized data formats and toolsets. Overall, the applications that drive work in BTSs are only a small subset of the usage that we anticipate for the Synchronic web.

### A.3. Distributed Ledger Technology

DLTs (Distributed Ledger Technologies) are decentralized networks that use blockchain technology to share a ledger between multiple distrusting parties. Examples of networks that operate in permissionless environments include Bitcoin [24] and Ethereum [31]. Examples of networks that operate in permissioned environments include Hyperledger Fabric [4] and Quorum [10]. In principle, most DLTs allow clients to commit data in a way that is, all things being equal (namely, the trustworthiness of nodes), at least as secure as the Synchronic Web.

DLTs are similarly secure and strictly more capable relative to the Synchronic Web. However, they are much less scalable for several reasons. First, the consensus protocol creates a bottleneck for transaction throughput. Permissionless blockchains like Bitcoin and Ethereum support on the order of ten transactions per second worldwide[29]. Modern permissioned blockchains like Hyperledger Fabric and Quorum may perform about two orders of magnitude better[30]. Second, DLTs must store data in perpetuity to maintain their security guarantees. The combination of these two fundamental scalability limitations implies that standard DLTs are incapable of supporting the large-scale use cases supported by the Synchronic Web. The DLT industry is investing significant time and effort into researching novel scaling mechanisms to augment standard DLTs including directed acyclic graphs, sharding, sidechains, and many other layer-2 protocols [20]. While these approaches succeed in improving scalability, many impose additional security costs and all impose additional complexity costs.

---

[26] https://opentimestamps.org/
[27] https://originstamp.com/
[28] https://tierion.com/chainpoint/
[29] https://www.blockchain.com/explorer
[30] https://hyperledger.github.io/caliper-benchmarks/fabric/performance/

# APPENDIX B. SECURITY MODEL

This section explores the cybersecurity model and best practices for using the Synchronic Web. To analyze the security of the system described in Section 3, we consider a global adversary $\mathcal{A}$ that can interact with any entity in the network by reading, blocking, and creating arbitrary messages. In addition, we allow $\mathcal{A}$ to compromise up to f notaries where $f < \frac{n}{3}$ and n is the total number of notaries. Since we do not assume that $\mathcal{A}$ has infinite computational power, we make the following assumptions about the primitives that we use in our system design:

1. The cryptographic hash function `Hash` is irreversible and collision resistant.
2. The cryptographic signature scheme `Sign` is verifiably deterministic and unforgeable.
3. The global consensus protocol can always guarantee consistency if $f < \frac{n}{3}$.
4. All communication messages are passed through a perfectly anonymous routing network.
5. All honest ledgers are perfectly secure and cannot be compromised.

Finally, we introduce an Oracle $\mathcal{O}$ which can infer the implied root of a commitment according to Algorithm 3 and compare it against the blockchain root published by the notary network. Given these assumptions, we argue that the only negative outcome that $\mathcal{A}$ can force on an otherwise well-functioning network is a temporary and non-discriminating denial-of-service attack. As a reference, the following notation will be used:

$\mathcal{A}$: A global adversary.

$\mathcal{O}$: The Oracle.

$\mathcal{C}$: A specific instance of content.

## B.1. Privacy

**Claim.** $\mathcal{A}$ cannot guess the contents of a ledger or link the owner of a ledger with their real-world identity with success probability $> \frac{1}{m} + \epsilon$ where m is the number of active ledgers and $\epsilon$ is a negligibly small number.

**Premises.** The supporting arguments are as follows:

1. Since we establish Assumption 5, $\mathcal{A}$ cannot obtain in-band information from the ledger.
2. Since we establish Assumption 4, $\mathcal{A}$ cannot obtain out-of-band information between the ledger and the notary.
3. Since the only in-band data between the ledger and the notary is the key from Equation 1 and the value from Equation 2, since both the key and the value are outputs of the `Hash` function, and since we establish Assumption 1, $\mathcal{A}$ cannot obtain data from the notary about the ledger.

**Caveats.** First, in the realistic situation that Assumption 5 fails, $\mathcal{A}$ will be able to read the contents of the ledger. Thus, software security for ledgers is an important consideration, although one that is out-of-scope for this theoretical discussion. Second, contrary to Assumption 4, no communication network is perfectly anonymous. Thus, users who require strong anonymity should use tools like Tor [15] or VPN to further obscure their routing.

## B.2. Immutability

**Claim.** If $\mathcal{A}$ creates a valid proof for some content $\mathcal{C}$ at some index $i$, $\mathcal{A}$ cannot modify $\mathcal{C}$ after block $i$ has been published by a majority of honest notaries without also invalidating the proof.

**Premises.** The supporting arguments are as follows:

1. Since Equation 2 computes $\mathcal{C}$ as a preimage of value, since Algorithm 1 computes value as a preimage of root, and since we establish Assumption 1, $\mathcal{O}$ will compute a different root if $\mathcal{A}$ attempts to provide a different input $\mathcal{C}'$.

2. Since we establish Assumption 3, $\mathcal{A}$ cannot change the value of root for a given index after it has been published.

**Caveats.** First, in the real world, $\mathcal{O}$ is some piece of verifier software that a more powerful adversary could conceivably compromise. Thus, software security for the verifier is an important consideration, although one that is out-of-scope for this theoretical discussion. Second, the probability that Assumption 3 fails for a well-operated network is small but non-zero. Thus, ledgers should monitor out-of-band signals for such faulty behavior, since it would indicate a catastrophic failure of the underlying infrastructure.

## B.3. Consistency

**Claim.** $\mathcal{A}$ cannot commit $\mathcal{C}$ and a different $\mathcal{C}'$ that are both valid for the same tuple $\langle$`index`, `public-key`, `path`$\rangle$.

**Premises.** The supporting arguments are as follows:

1. Since we establish Assumption 2 and since Equation 1 computes key as a deterministic function of the tuple $\langle$`index`, `public-key`, `path`$\rangle$, and no other independent variables, $\mathcal{O}$ will always recompute the same key when given the same input tuple.

2. Since Algorithm 1 outputs a verifiable map containing unique keys and since Algorithm 2 only outputs the same proof for any single key, $\mathcal{A}$ can only produce a single valid local tuple $\langle$`local content`, `index`, `public-key`, `local path`$\rangle$ for which Algorithm 3 executed by $\mathcal{O}$ will return the correct local root.

3. Since Equation 2 computes the global value as a deterministic function of the local root and no other independent variables, $\mathcal{O}$ will always recompute the same global value when given the same local root.

4. Since Algorithm 1 outputs a verifiable map containing unique keys, and since Algorithm 2 only outputs the same proof for any single key, $\mathcal{A}$ can only produce a single valid global tuple ⟨global `content`, `index`, `public-key`, global `path`⟩ for which the Algorithm 3 executed by $\mathcal{O}$ will return the correct global root.

**Caveats.** See caveats in Section B.2. In addition, the task of converging onto well-known, semantically meaningful paths is a vital but nontrivial step toward leveraging the practical value of consistency guarantees.

### B.4. Denial-of-Service

**Claim** Denial-of-service attacks by $\mathcal{A}$ cannot last longer than $b$ blocks or target any honest ledger with success probability $> \frac{1}{m} + \epsilon$ where $b$ is the parameter defined in Section 3.2.3 under the weak availability guarantee, $m$ is the number of active ledgers, and $\epsilon$ is a negligibly small number.

**Premises.** The supporting arguments are as follows:

1. Since we establish Assumption 3, if $\mathcal{A}$ compromises the leader notaries, then the only block for which it can deny service is the current block that it is allowed to publish.

2. Since we establish the privacy of honest ledgers in Section B.1, $\mathcal{A}$ cannot target any particular honest ledger and can only deny service indiscriminately.

3. Since the consensus requirements described in Section 3.2.3 mandate the detection and removal of uncooperative leader notaries within $b$ blocks, $\mathcal{A}$ can only maintain a sustained denial-of-service attack for up to $b$ blocks.

**Caveats.** Since the combined probability of a malicious denial-of-service attack by $\mathcal{A}$ and incidental network interruptions for honest ledgers is non-zero, verifiers should implement logic at the application level to tolerate a logically acceptable number of missing proofs by the ledger. For instance, consider the situation in which a verifier needs to confirm that a ledger that operates an immutable database has not forked the database within the past $j$ blocks. In this situation, the verifier only requires $\geq \frac{j}{2} + 1$ proofs to confirm the absence of a fork, thereby tolerating up to $\frac{j}{2} - 1$ missing proofs.

## NOMENCLATURE

**authenticator**  An object used to assert the control of a cryptographic identity over a commitment.

**commitment**  An object asserting the provenance of a piece of content in the Synchronic Web.

**content**  A piece of data that is committed to the Synchronic Web.

**current**  The index of the next upcoming unpublished block.

**index**  An incrementally increasing number that uniquely identifies a block.

**key**  A string that defines the location at which a value is stored in a verifiable map.

**ledger**  A client of the Synchronic Web that writes commitments.

**linker**  An object used to connect multiple commitments in a semantically meaningful way.

**notary**  A server of the core Synchronic Web interface.

**path**  The portion of a commitment that identities the semantic context of the content.

**periodicity**  An integer that determines the frequency at which a ledger commits content.

**promise**  A response by a notary asserting its intent to fulfill a commitment request.

**proof**  The portion of a commitment that validates the connection between the value and the root.

**resolver**  An object used to retrieve hashable content from a target data store.

**root**  The topmost node of a verifiable map.

**sequence**  An incrementally increasing number that identifies the version of a piece of content.

**value**  The string that is stored in a verifiable map at a specific key.

**verifiable map**  A map that facilitates proofs of inclusion for key-value pairs.

**verifier**  A client of the Synchronic Web that reads commitments.

# ERRATA

| Location | Correction |
|---|---|
| Algorithm 3 | Modify function header and section to include the omitted value parameter. |
| Section 3.2.2 | Specify that Alice should include the index into her content, not path. |
| Figure 3-2 | Modify the bottom-left box to simplify and clarify the definition of $k$ (the key) and $v$ (the value). |
| Section A | Specify that signatures are verifiably deterministic rather than just deterministic. |
| All | Change terminology from journal to ledger. |